\documentclass[10pt,english]{article}
\usepackage[T1]{fontenc}
\usepackage{babel}
\usepackage{graphics}
\usepackage{psfig}
\usepackage{setspace}
\makeatletter

\providecommand{\LyX}{L\kern-.1667em\lower.25em\hbox{Y}\kern-.125emX\@}
\newcommand{\noun}[1]{\textsc{#1}}

 \usepackage{verbatim}
 \newcommand{\lyxaddress}[1]{
   \par {\raggedright #1 
   \vspace{1.4em}
   \noindent\par}
 }

\usepackage{vmargin}
\usepackage{mathrsfs}
\usepackage{amssymb}
\usepackage{bbold}
\newcommand{\feyn}[1]{{#1}\!\!\!{\slash}}

\newcommand{\beq}{\begin{equation}}
\newcommand{\eeq}{\end{equation}}
\newcommand{\LL}{\mathscr{L}}

\newcommand{\llt}{\langle \! \langle}
\newcommand{\rrt}{ \rangle \! \rangle}

\newcommand{\uc}{ u^{\dagger}}
\newcommand{\dc}{ d^{\dagger}}
\newcommand{\Sc}{ s^{\dagger}}
\newcommand{\ds}{\Delta_{qs}}
\newcommand{\dq}{\Delta_{qq}}
\newcommand{\dNdq}{\frac{\partial N}{\partial \ll \bar{q}q \gg}}
\newcommand{\dNds}{\frac{\partial N}{\partial \ll \bar{s}s \gg}}

\newcommand{\dEda}{\frac{\partial E^-}{\partial \alpha}}
\newcommand{\dEsda}{\frac{\partial E^-_s}{\partial  \alpha }}
\newcommand{\dNda}{\frac{\partial N(p)}{\partial \alpha }}

\newcommand{\dxda}[1]{\frac{\partial {#1}}{\partial \alpha }}

\makeatother
\date{}
\begin{document}

\title{Thermodynamics of the 3-flavor NJL model : chiral symmetry breaking and color
superconductivity}

\author{ F. Gastineau \protect\( ^{1}\protect \),R. Nebauer \protect\( ^{1,2}\protect \) and J.Aichelin \protect\( ^{1}\protect \) }

\maketitle

\lyxaddress{\centering \textit{\small \protect\( ^{1}\protect \) SUBATECH, Laboratoire
EMN, IN2P3-CNRS et Université de Nantes, }\\
\textit{\small F-44072 Nantes Cedex 03, France }\\
\textit{\small \protect\( ^{2}\protect \)Institute for Theoretical Physics
Universität Rostock, Rostock, Germany}\small }

\begin{abstract}
Employing an extended three flavor version of the NJL model we discuss in detail
the phase diagram of quark matter. The presence of quark as well as of diquark
condensates gives rise to a rich structure of the phase diagram. We study in
detail the chiral phase transition and the color superconductivity as well as
color flavor locking as a function of the temperature and chemical potentials
of the system. 
\end{abstract}
\section*{Introduction}

At low temperatures and densities all quarks are confined  into hadrons. In
this phase the chiral symmetry is spontaneously broken by the quark condensates.
Raising the temperature, one expects that the chiral symmetry becomes restored
and the quarks are free. This state is called quark gluon plasma (QGP). In the
QGP all symmetries of the QCD Lagrangian are restored. For QCD at low temperatures
and high densities one expects a phase where the quarks are in a superconducting
state \cite{new1,new2,alf98,new3}. All this different phases define the phase diagram of QCD 
\cite{raja99}
in the plane of temperature and density. This phase diagram is not directly
accessible. QCD calculations are only possible on a lattice at zero baryon density.
In order to explore the finite temperature and density region, one has to rely
on effective models.  Two types of such effective models has been advanced
to study the high density low temperature section 
The first type of models includes weak coupling QCD calculations, including the
gluon propagators \cite{new4}. The second type include instanton 
\cite{new3,berg99,new5} as well as Nambu and Jona-Lasinio (NJL) models \cite
{new2,alf98a}. 
These models shows a color
superconducting phase at high density and low temperature. In this phase the
\( SU_{C}(3) \) color symmetry of QCD breaks down to an \( SU_{C}(2) \). Including
a third flavor, another phase occurs, the color-flavor-locked state (CFL) of
quark matter \cite{alf98a}\cite{raja98b}\cite{sch99}. 

The two flavor results of the instanton approach are reproduced by the model
of Nambu and Jona-Lasinio (NJL) \cite{njl1}\cite{njl2} if one includes an
appropriate interaction as was shown by Schwarz et al. \cite{skp99}. This model
has been extended by Langfeld and Rho \cite{lan98} who included all possible
interaction channels and discovered an even richer phase structure of the QCD
phase diagram, including a phase where Lorentz symmetry is spontaneously broken.

The choice of the NJL model is motivated by the fact that this model displays
the same symmetries as QCD and that it describes correctly the spontaneous breakdown
of chiral symmetry in the vacuum and its restoration at high temperature and
density. In addition, the NJL model has been successfully used to describe the
meson spectra and thus is able to reproduce the low temperature and low density
phenomena of QCD \cite{hats94}\cite{klev92}\cite{rip}. Thus, this is a model
which starts out from the opposite direction as the instanton model which is
motivated as high density approximation of QCD. Therefore it is interesting
to see whether the NJL model is able to describe the other phases, the color
superconducting phase and the color flavor locking observed in the instanton
approach.  The shortcoming of the NJL model is the fact that it does not describe
confinement, or more generally any gauge dynamics at all. 
Here we will evaluate the thermodynamical properties of the quarks
in the NJL model at finite temperature and density and we will discuss the symmetries
of the  different phases. We present numerical results for the calculation of
the different condensates. For our study of the phase diagram we use one specific
set of parameters.We treat the three flavor version of the model, including
an interaction in the quark-antiquark channel, a t'Hooft interaction and an
interaction in the diquark channel. We restrict ourself to the scalar/ pseudoscalar
sector of these interactions. 

The paper is organized as follows: In chapter \ref{chap_model} we will briefly
review the NJL model and present the Lagrangian we will use. In chapter \ref{chap_chiral}
we study the quark condensate and the restoration of chiral symmetry.  In chapter
\ref{chap_super} we add the interaction in the diquark channel and present
the numerical results for the color superconducting sector. We will have a complete
evaluation of the phase diagram of the NJL model, including chiral and superconducting
phase transition at finite temperature and (strange and light quark) density.
In chapter \ref{chap_concl} we present our conclusions.

\section{The model \label{chap_model}}

The model we  use is an extended version of the NJL model, including an
interaction in the diquark channel. In fact, the NJL model can be shown to be
the simplest low energy approximation of QCD. It describes the interaction between
two quark currents as a point-like exchange of a perturbative gluon \cite{dhar84}\cite{ebe94}.
Applying an appropriate Fierz-transformation to this interaction, the Lagrangian
separates into two pieces: a color singlet interaction between a quark and an
antiquark (\( \LL _{(\bar{q}q)} \)) and a color antitriplet interaction between
two quarks \( \LL _{(qq)} \). The color singlet channel is attractive in the
scalar and pseudoscalar sector and repulsive in the vector and pseudovector
channel. The Lagrangian in the diquark sector has two parts, both attractive:
a flavor antisymmetric and flavor symmetric channel. The former includes Lorentz
scalar, pseudoscalar and vector interactions, the latter a pseudo scalar interaction
only. 

The coupling constants of these different channels are related to each other
by the Fierz transformation. Due to the extreme simplification of the gluon
propagator in this approximation, the resulting model cannot reproduce confinement
which is described by the infrared behavior of the gluon propagator. 

The resulting Lagrangian has a global axial symmetry \( U_{A}(1) \), and an
extra term \( \LL _{A} \) in the form of the t'Hooft determinant is added in
order to break explicitly this symmetry. The resulting Lagrangian has then the
general form:

\begin{equation}
\LL =\LL _{0}+\LL _{(\bar{q}q)}+\LL _{(qq)}+\LL _{A}
\end{equation}
where \( \LL _{0} \) is the free kinetic part.

The interaction part of the Lagrangian has a global color, flavor and chiral
symmetry. Chiral symmetry is explicitly broken by non zero current quark masses,
flavor symmetry by a mass difference between the flavors.

The different interaction channels of this Lagrangian give rise to a very rich
structure of the phase diagram, which was completely evaluated in the two flavor
case by Langfeld and Rho \cite{lan98}. Here we will concentrate on the three
flavor case. The evaluation of the complete phase structure in the three flavor
case is a quite difficult task and we will concentrate here on the Lorentz scalar
and pseudoscalar interactions. In the mesonic channel this interaction is responsible
for the appearance of a quark condensate and for the spontaneous breakdown of
the chiral symmetry. In the diquark channel it gives rise to a diquark condensate
which can be identified with a superconducting gap. 

Describing the quark fields by the Dirac-spinors \( q \), the Lagrangian we
will use here has the form:

\begin{eqnarray}
\LL  & = & \bar{q}(i\feyn {\partial }-m_{0f})q+G_{S}\sum ^{8}_{a=0}\left[ (\bar{q}\lambda _{F}^{a}q)^{2}+(\bar{q}i\gamma _{5}\lambda _{F}^{a}q)^{2}\right] \label{Lagr_allgemein} \\
 &  & +G_{DIQ}\sum ^{3}_{k=1}\sum ^{3}_{\gamma =1}\left[ (\bar{q}_{i,\alpha }\epsilon ^{ijk}\epsilon ^{\alpha \beta \gamma }q_{j,\beta }^{C})(\bar{q}_{i',\alpha '}^{C}\epsilon ^{i'j'k}\epsilon ^{\alpha '\beta '\gamma }q_{j',\beta '})\right] \nonumber \\
 &  & +G_{DIQ}\sum ^{3}_{k=1}\sum ^{3}_{\gamma =1}\left[ (\bar{q}_{i,\alpha }i\gamma _{5}\epsilon ^{ijk}\epsilon ^{\alpha \beta \gamma }q_{j,\beta }^{C})(\bar{q}_{i',\alpha }^{C}i\gamma _{5}\epsilon ^{i'j'k}\epsilon ^{\alpha '\beta '\gamma }q_{j',\beta '})\right] \nonumber \\
 &  & +G_{D}\left[ det\bar{q}(1-i\gamma _{5})q+det\bar{q}(1+i\gamma _{5})q\right] .\nonumber 
\end{eqnarray}
The first term is the free kinetic part, including the flavor dependent current
quark masses \( m_{0f} \) which break explicitly the chiral symmetry of the
Lagrangian. The second part is the scalar/ pseudoscalar interaction in the mesonic
channel, it is diagonal in color. The matrices \( \lambda _{F} \) act in the
flavor space. The third part describes the interaction in the scalar/ pseudoscalar
diquark channel. The charge conjugated quark fields are denoted by \( q^{C}=C\bar{q}^{T} \)
and the color (\( \alpha ,\beta ,\gamma  \)) and flavor (\( i,j,k \)) indices
are displayed explicitly. We note that due to the charge conjugation operation
the product \( \bar{q}i\gamma _{5}q^{C} \) is a Lorentz scalar. This interaction
is antisymmetric in flavor and color, expressed by the completely antisymmetric
tensor \( \epsilon ^{ijk} \). Finally we add the six point interaction in the
form of the t'Hooft determinant which breaks explicitly the \( U_{A}(1) \)
symmetry of the Lagrangian. The \( det \) runs over the flavor degrees of freedom,
consequently the flavors become connected.

The NJL model is non renormalizable, thus it is not defined until a regularization
procedure has been specified. As we are interested in the thermodynamical properties
of the model, calculated with help of the thermodynamical potential, we will
use a three dimensional cut-off in momentum space. This cut-off limits the validity
of the model to momenta well below the cut-off.

The model contains six parameters: The current mass of the light and strange
quarks, the coupling constants \( G_{D} \) and \( G_{S} \) and the momentum
cut-off \( \Lambda  \) which are fixed by physical observables: the pion and
kaon mass, the pion decay constant, the mass difference between \( \eta  \)
and \( \eta ' \) , once the mass of the light quarks was fixed, as well as by
the vacuum value of the condensate $<q\bar q>^{1/3} = - 230 MeV$. The last parameter
is the coupling constant in the diquark channel \( G_{DIQ} \). For the mesonic
sector we will use the parameters of \cite{rehb}: a current light quark mass
\( m_{0q}=5.5\, MeV \), a current strange quark mass \( m_{0s}=140.7\, MeV \),
a three dimensional ultraviolet cut-off \( \Lambda =620\, MeV \), a scalar
coupling constant \( G_{S}=1.835/\Lambda ^{2} \) and a determinant coupling
\( G_{D}=12.36/\Lambda ^{5} \).. This parameter set results in effective vacuum quark masses 
of \( m_{q}=367.6\, MeV \)
, \( m_{s}=549.5\, MeV \) and the quark condensates are
 \( \llt \bar{q}q\rrt =(-242\, MeV)^{-3} \)
and \( \llt \bar{s}s\rrt =(-258\, MeV)^{-3} \). 

We perform our calculations in the mean field approach for an operator product

\begin{equation}
\hat{\rho }_{1}\hat{\rho }_{2}\approx \hat{\rho }_{1}\llt \rho _{2}\rrt +\llt \rho _{1}\rrt \hat{\rho }_{2}-\llt \rho _{1}\rrt \llt \rho _{2}\rrt 
\end{equation}
 where \( \llt \rho \rrt  \) is the thermodynamical average of the operator
and the fluctuations around this mean value are supposed to be small. We will
apply this approximation to the products of quark fields appearing in the interaction
part of the Lagrangian.

\section{Chiral phase transition\label{chap_chiral}}

We start our study with an investigation of the quark-antiquark sector and the
chiral phase transition. The diquark sector is subject of the next section. 

The NJL model displays the right features of the chiral symmetry breaking. On
the one hand we have an explicitly broken chiral symmetry by the inclusion of
a small current quark mass. On the other hand, the model describes correctly
the spontaneous breakdown of chiral symmetry: the existence of a quark condensate,
responsible for a high effective quark mass and the existence of massless (or
very light, if the chiral symmetry is explicitly broken) Nambu-Goldstone bosons.
Lattice QCD calculations show that at a temperature of \( \approx 170MeV \)
the chiral symmetry is restored (the quark condensates melt at increasing temperature),
a result which is reproduced by the NJL model \cite{hats94}\cite{klev92}\cite{asa89}.
As the region of finite density is not accessible to lattice QCD calculations,
the chiral phase transition at high density is a subject of speculation. The
point and the order of the chiral phase transition in the temperature - density
plane defines the phase diagram. Here we will present such a phase diagram for
the three flavor NJL model and a specified set of parameters. This phase diagram
can be viewed as an approximation of the QCD phase diagram, but we have to take
into account that the NJL model does not describe confinement (we always have
a gas of quarks and not a gas of hadrons) and that the degrees of freedom are
not the same as in QCD (the model contains no gluons). Here we will focus on
the thermodynamical properties of the quarks described in the color singlet
channel of the Lagrangian (\ref{Lagr_allgemein}), this means the thermodynamical
properties of the quark condensates and masses.

For the study of the thermodynamical properties of the quark-antiquark sector
we will evaluate the thermodynamical potential in the mean field approximation.
We start out from the Lagrangian in the mean field approximation
\begin{equation}
\LL ^{MF}=\bar{q}(i\feyn {\partial }-M)q-2G_{S}
(\alpha ^{2}+\beta ^{2}+\gamma ^{2})+4G_{D}\alpha \beta \gamma .
\end{equation}
 where \( M_{f} \) is the effective quark mass (defined via the quark condensates
\( \llt \bar{q}q\rrt  \))

\begin{eqnarray}
M_{f} & = & m_{0f}-4G_{S}\llt \bar{q}_{f}q_{f}\rrt +2G_{D}\llt \bar{q}_{f_{1}}q_{f_{1}}\rrt \llt \bar{q}_{f_{2}}q_{f_{2}}\rrt \quad f\neq f_{1}\neq f_{2}\label{gap_intr} \\
 & = & m_{0f}+\delta m_{f}\nonumber 
\end{eqnarray}
 and the quark condensates are written in a short hand notation
\begin{equation}
\alpha =\llt \bar{u}u\rrt \quad \beta =\llt \bar{d}d\rrt \quad \gamma =\llt \bar{s}s\rrt .
\end{equation}
 The mean field Hamiltonian

\begin{equation}
H^{MF}=\int d^{3}x\sum _{f=\{u,d,s\}}\left[ \bar{q}_{f}i\gamma ^{0}\partial _{0}q_{f}+2G_{S}(\alpha ^{2}+\beta ^{2}+\gamma ^{2})-4G_{D}\alpha \beta \gamma \right] 
\end{equation}
 is transformed into an operator \( \hat{H} \) in second quantization using
\begin{equation}
\label{spinorop}
\hat{q}_{f}(x)=\sum _{s=\pm }\int \frac{d^{3}p}{(2\pi )^{3}}\left[ \hat{a}_{\vec{p},s,f}u_{f}(p,s)e^{-ipx}+\hat{b}_{\vec{p},s,f}^{\dagger }v_{f}(p,s)e^{ipx}\right] .
\end{equation}
 At the moment, the quark condensates are unknown quantities. In order to evaluate
them, we calculate the grand-canonical potential 
\begin{equation}
\Omega =-\frac{1}{\beta }Tr\left[ e^{-\beta (\hat{H}-\mu \hat{N})}\right] 
\end{equation}
 with \( \mu  \) being the chemical potential, \( \beta  \) the inverse temperature
and \( \hat{N} \) the particle number operator:

\begin{equation}
\hat{N}=\hat{n}(\vec{p},s,f,c)-\hat{\bar{n}}(\vec{p},s,f,c)
\end{equation}
 where \( \hat{n}(\vec{p},s,f,c) \) ,\( \hat{\bar{n}}(\vec{p},s,f,c) \) are
the number operators for particles and antiparticles with momentum \( \vec{p} \),
spin \( s \), flavor \( f \) and color \( c \). These operators are defined
via the creation and annihilation operators for particles \( \hat{n}(\vec{p},s,f,c)=\hat{a}_{\vec{p},s,f,c}^{\dagger }\hat{a}_{\vec{p},s,f,c} \)
and antiparticles \( \hat{\bar{n}}(\vec{p},s,f,c)=\hat{b}_{\vec{p},s,f,c}^{\dagger }\hat{b}_{\vec{p},s,f,c} \).
We consider the condensates as parameters with respect to which the potential
has to be minimized. The appearance of the quark condensates breaks spontaneously
the chiral symmetry of the original Lagrangian. 

In second quantization the exponent of the chemical potential reads as follows
: 
\begin{eqnarray}
(\hat{H}^{MF}-\mu \hat{N})/V & = & \sum _{s,f,c}\int ^{\Lambda }_{0}\frac{p^{2}dp}{2\pi ^{2}}\left[ E_{\vec{p},f}-(E_{\vec{p},f}-\mu _{f})\hat{n}(\vec{p},s,f,c)\right. \nonumber \\
 &  & \left. -(E_{\vec{p},f}+\mu _{f})\hat{\bar{n}}(\vec{p},s,f,c)\right] \nonumber \\
 &  & +\left[ 2G_{S}(\alpha ^{2}+\beta ^{2}+\gamma ^{2})-4G_{D}\alpha \beta \gamma \right] 
\end{eqnarray}
 where \( V \) denotes the volume we have integrated out. The energy \( E_{\vec{p},f}=\sqrt{M^{2}_{f}+\vec{p}^{2}} \)
depends on the flavor of the quarks and their momentum but is independent of
color or spin. The evaluation of the grand canonical potential in the mean field
approximation gives the result:
\begin{eqnarray}
\frac{\Omega ^{MF}}{V} & = & 2G_{S}(\alpha ^{2}+\beta ^{2}+\gamma ^{2})-4G_{D}\alpha \beta \gamma \nonumber \\
 &  & -\frac{N_{c}}{\pi ^{2}}\sum _{f=\{u,d,s\}}\int _{0}^{\Lambda }p^{2}dp\left\{ E_{\vec{p},f}+\frac{1}{\beta }ln[1+e^{-\beta (E_{\vec{p},f}-\mu _{f})}]\right. \label{thpot} \\
 &  & \left. +\frac{1}{\beta }ln[1+e^{-\beta (E_{\vec{p},f}+\mu _{f})}]\right\} .\nonumber 
\end{eqnarray}
It has  to be minimized with respect to the quark condensates :

\begin{equation}
\frac{\partial \Omega ^{MF}}{\partial \llt \bar{q}_{f}q_{f}\rrt }=0.
\end{equation}
 We obtain three equations, one for each quark condensate
\begin{equation}
\llt \bar{q}_{f}q_{f}\rrt =-M_{f}\frac{N_{C}}{\pi ^{2}}\int ^{\Lambda }_{0}dp\frac{p^{2}}{E_{\vec{p}}}\left[ 1-f(E_{\vec{p},f}+\mu _{f})-f(E_{\vec{p},f}-\mu _{f})\right] 
\end{equation}
where we defined the Fermi function \( f(x)=(1+exp(-\beta x))^{-1} \). The
equations for the quark condensates are coupled (see eq.\ref{gap_intr}). For three
flavors we have thus three coupled gap equations which have to be solved self-consistently.
Their solution, displayed in appendix B, enables us to calculate the quark condensates and quark masses
at finite temperature and chemical potential (density). 

We have to take care about the limits of the theory: The regularization cut-off
of the theory implies that the chemical potential has always to be smaller than
this cut-off and that the temperature has not to be too elevated: The Fermi
function will be smoothly extended to high momenta and we have to take into
account that all states above the cut-off are ignored by the model. 

The condensate is responsible for the spontaneous breakdown of chiral symmetry
at low densities and temperatures. At high temperature and density the quark
condensate drops (it becomes very small, or zero - in the case of zero current
quark masses) and consequently chiral symmetry is restored (up to the current
quark masses). Hence the quark condensate is the order parameter of the chiral
phase transition. The phase transitions we are dealing with are - depending
on the parameters and of the density respective temperature - of first or second
order or of the so called cross-over type and we can classify the phase transition
by means of this order parameter. The first order phase transition is specified
by a discontinuity in the order parameter, for the second order phase transition
the order parameter is continuous but not analytical at the point of the phase
transition. The third type, the cross-over, is not a phase transition in the
proper sense. Here the order parameter does not display a non-analytical point
but it shows a smooth behavior. 

\begin{figure}
{\par\centering \resizebox*{1\textwidth}{!}{\includegraphics{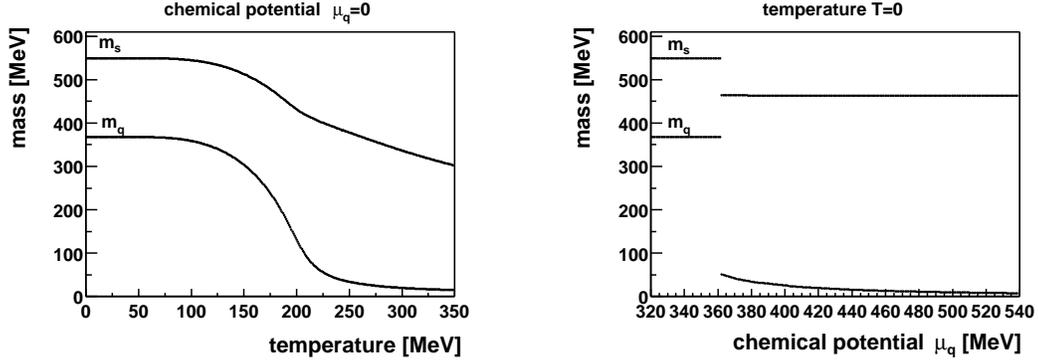}} \par}

\caption{\emph{The mass of strange and light quarks as a function of the temperature
(lhs) and as a function of the light quark chemical potential (rhs)} \label{bild_massqs_t_muq}}
\end{figure}

In a first step we will consider the chiral phase transition as a function of
temperature and chemical potential of the light quarks, the strange quark density
is supposed to be zero. In figure~\ref{bild_massqs_t_muq}, lhs, we plot the
mass of the light and strange quarks as a function of temperature at zero baryon
density for the parameters presented above. 

At zero density we observe a smooth cross-over of the chiral phase transition
as a function of temperature: at low temperature the chiral symmetry is spontaneously
broken, with rising temperature the quark condensate melts away and the quark
masses approach the current mass, at least for the light quarks. For the strange
quarks we observe a much smoother transition and at the highest temperature
we can treat in the framework of the NJL model (approximately \( 230\, MeV \))
their mass is still higher than their current mass . This smooth cross-over
we observe only for the special case of three non-zero current quark masses. 

At zero temperature we observe for our parameter set a first order phase transition.
As a function of the chemical potential the light quark mass drops suddenly
to a value close to the current quark mass. The strange quarks change slightly
their mass due to the coupling between the flavors. For higher values of the
chemical potential the strange quark mass is stable. The light quark condensate
is too small for a change of the strange quark mass. Only a rise of the chemical
potential of the strange quarks can drop the strange quark mass further, as
will be discussed in the last part of this section where we present the extension
to strange quark matter.

\begin{figure}
{\par\centering \resizebox*{1\textwidth}{!}{\includegraphics{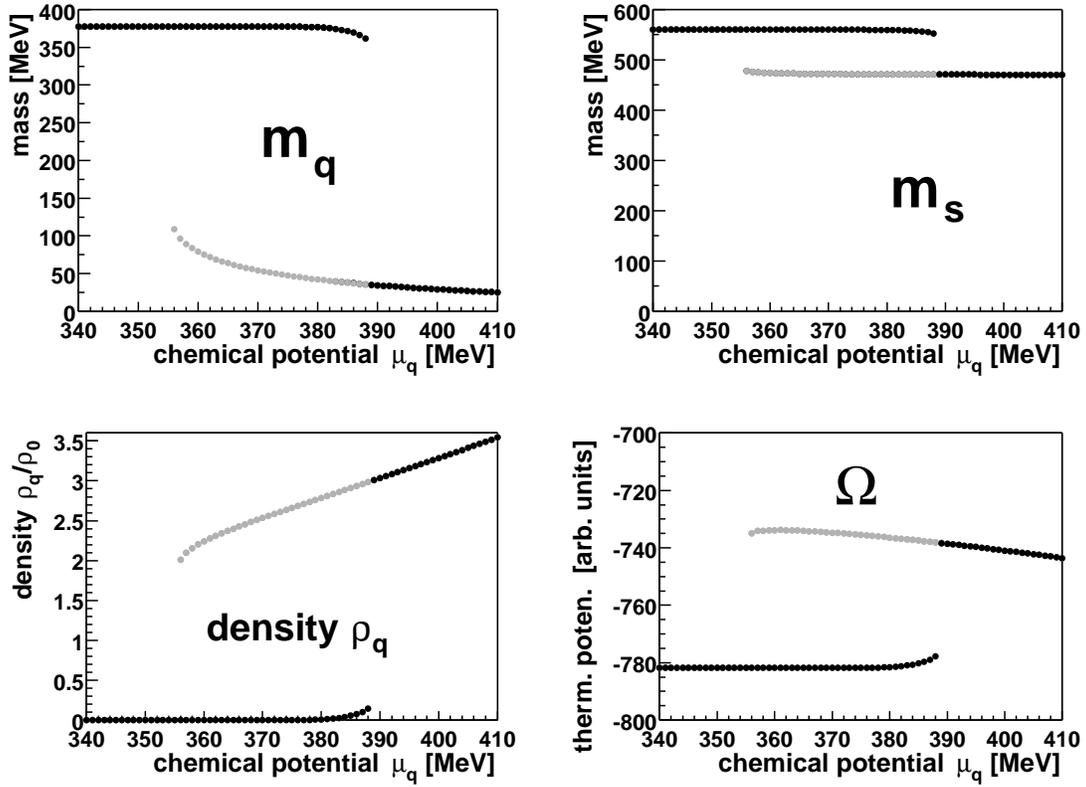}} \par}

\caption{\emph{Detailed representation of the first order phase transition as a function
of the light quark potential at zero temperature. On the top: the light (lhs)
and strange (rhs) quark masses bottom: the density (lhs) and the thermodynamical
potential (rhs). The light lines represent the metastable region, the dark lines
the stable region (minimization of the thermodynamical potential).}\label{bild_detailphase_muq}}
\end{figure}

A first order phase transition is characterized by the existence of metastable
phases, the equivalent of for example oversaturated vapor. These metastable
phases are a solution of the gap equation, but their thermodynamical potential
is larger than for the stable phase. We show this in detail in figure~\ref{bild_detailphase_muq}.
On the top we display the quark mass (light and strange), on the bottom the
density of light quarks and the thermodynamical potential. The stable phases
which minimize the thermodynamical potential are shown as dark lines, the metastable
phases as light lines. 

For the mass of the light quarks we observe the transition from the stable phase
at high chemical potential to a state whose mass is larger than its chemical
potential, this means to zero density. Increasing the chemical potential yields
a first order phase transition, i.e. the mass of the quarks drops suddenly.
This abrupt change in the quark masses gives rise to a jump in the density -
for a constant chemical potential suddenly much more states become accessible.
This implies at the same time that certain densities does not exist. 
In our case the normal nuclear matter density is just in this region and there
are nice explications for this fact \cite{bulb96}\cite{alf98}. For the interpretation
one has to remember that we are talking about a quark  gas without confinement.
Here, nuclear matter at normal density one has to consider as a phase which
contains dense droplets of quarks in which chiral symmetry is restored, surrounded
by the vacuum or a very diluted quark gas (which should be confined in QCD).
The size of these droplets is not given by the theory, but it is not farfetched
to identify these objects with the nucleons. 

\begin{figure}
{\par\centering \resizebox*{1\textwidth}{!}{\includegraphics{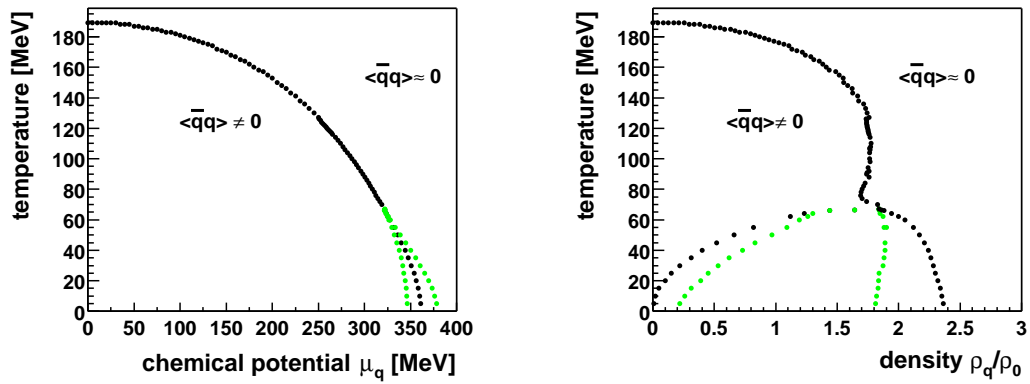}} \par}

\caption{\emph{Phase diagram for the mass of the light quarks (chiral phase transition)
as a function of temperature and the light quark chemical potential (lhs) and
density (rhs). The dark lines represent the transition by the stable phases,
the light lines the transition by metastable phases in case of a first order
phase transition.} \label{bild_ptrans_tmuq}}
\end{figure}

We observe thus for our set of parameters a first order phase transition as
a function of the chemical potential at zero temperature and cross-over as a
function of temperature at zero density. Extrapolating now to the plane of finite
temperature and chemical potential there must be a point where both kinds of
phase transition join, the so called tricritical point. In figure~\ref{bild_ptrans_tmuq}
we show this phase diagram at finite temperatures and chemical potentials (lhs,
at the rhs as a function of density). We display as dark lines the transition
by the stable state (or the transition line for the cross-over) and as light
lines the metastable phases. The tricritical point is located at a temperature
\( T=66\, MeV \) and a chemical potential of \( \mu _{q}=321\, MeV \) which
corresponds to a density of \( \rho _{q}=1.88\rho _{0} \). 
The location of the tricritical point depends strongly on the choice of the
cut-off and of the coupling constant \cite{pol}. 

\begin{figure}
{\par\centering \resizebox*{1\textwidth}{!}{\includegraphics{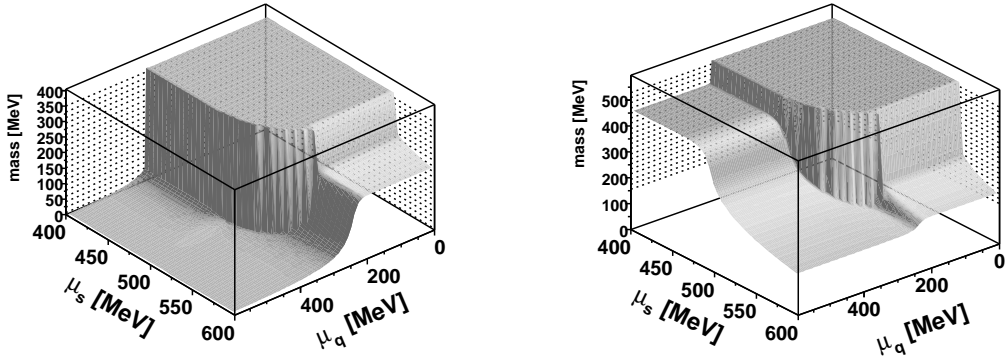}} \par}

\caption{\emph{Light (lhs) and strange (rhs) quark masses as a function of the chemical
potential of light (\protect\( \mu _{q}\protect \)) and strange (\protect\( \mu _{s}\protect \))
quarks at zero temperature.} \label{bild_mqms_muqs}}
\end{figure}

In figure~\ref{bild_mqms_muqs} we plot the quark masses (light, lhs and strange,
rhs) as a function of the chemical potential of light and strange quarks at
zero temperature. We can see the influence of the coupling between the flavors
as already discussed for the light quark chemical potential. The strange quark
mass drops suddenly at high chemical potentials of the strange quarks \( \mu _{s} \)
and low chemical potentials for the light quarks \( \mu _{q} \). Once the chiral
phase transition for the light quarks has taken place (at high values of \( \mu _{q} \)),
the strange quark mass shows a cross-over transition for high \( \mu _{s} \).
For high values of \( \mu _{q} \) and \( \mu _{s} \), both quark masses have
a value close to their current quark mass.
With increasing temperature, the phase transitions will take place at lower
values of the chemical potentials. This is very pronounced for the light quarks
(see figure~\ref{bild_massqs_t_muq}) and less for the strange quarks which
change their mass quite slowly with temperature due to the high current quark
mass (compare figure~\ref{bild_massqs_t_muq}).

\section{Color superconductivity\label{chap_super}}

In this section we will study the diquark channel. We will see that quarks which
have opposite spin and momenta condense in the scalar channel into diquarks.
This resembles superconductivity \cite{schrieff}\cite{fewa}. Here we have
in addition a complex structure in color and flavor space. In classical superconductivity
the condensation occurs close to the Fermi surface. In our case we have to take
into account that quarks with different flavors may have different Fermi-surfaces.
Because the coupling between the quarks is quite small, the condensation will
only occur if the Fermi momenta of the two quarks are quite close to each other. 

In order to calculate the properties of the NJL model in the superconducting
sector, we will apply the generalized thermodynamical approach of the Hartree-Bogolyubov
theory to quark matter (see for example \cite{iwa95}) described by the Lagrangian
(\ref{Lagr_allgemein}).

The Lagrangian (\ref{Lagr_allgemein}) in the mean field approximation including
the diquark sector reads as follows:

\begin{eqnarray}
\LL ^{MF} & = & \sum \bar{q}(i\feyn {\partial }-M)q-2G_{S}(\alpha ^{2}+\beta ^{2}+\gamma ^{2})+4G_{D}\alpha \beta \gamma \nonumber \\
 &  & +\bar{q}_{i\alpha }\frac{\tilde{\Delta }^{k\gamma }}{2}q_{j\beta }^{C}+\bar{q}_{i\alpha }^{C}\frac{\tilde{\Delta }^{k\gamma \dagger }}{2}q_{j\beta }-\sum _{k,\gamma }\frac{|\Delta ^{k\gamma }|^{2}}{ 4G_{DIQ}}\label{lag_su3_diq} 
\end{eqnarray}
 greek indices denote the colors, latin indices the flavors. 

The diquark condensate is defined by:

\begin{equation}
\label{diq_su3}
\tilde{\Delta }^{k\gamma }=2G_{DIQ}i\gamma _{5}\epsilon ^{\alpha \beta \gamma }\epsilon ^{ijk}\llt \bar{q}_{i'\alpha '}i\gamma _{5}\epsilon ^{i'j'k}\epsilon ^{\alpha '\beta '\gamma }q_{j'\beta '}^{C}\rrt =i\gamma _{5}\epsilon ^{\alpha \beta \gamma }\epsilon ^{ijk}\Delta ^{k\gamma }
\end{equation}
This diquark condensate occurs for all three colors simultaneously. We note that as in classical superconductivity
the baryon (or particle) number is not conserved. Hence the electromagnetic
\( U_{em}(1) \) symmetry is spontaneously broken and Goldstone bosons appear
in the form of Cooper pairs. The diquark condensate carries a color and a flavor
index. For a given flavor and color the condensate is completely antisymmetric
in the two other flavors and colors. The condensate \( \Delta ^{sr} \) is created
for example by green and blue up and down quarks. 

The diquark condensate is completely antisymmetric in the color degrees of freedom,
a property which is only shared by three of the eight Gell Mann matrices which
generate the \( SU_{C}(3) \). Hence a  finite diquark condensate breaks down
the \( SU_{C}(3) \) color symmetry to a \( SU_{C}(2) \) if the mass of the strange quark is heavy. The same is true
for the flavor sector if the three flavors are degenerated in mass. For two
flavors only the Lagrangian is invariant with respect to a chiral transformation.
If the diquark condensates coexist for all three flavors, the chiral symmetry
is spontaneously broken. 

Due to the product of two antisymmetric tensors the symmetry is even more reduced
if all three quark flavors form a diquark condensate. In order to see this,
we assume first that all three colors (for one flavor) are equivalent. Than
we can assume without loss of generality that \( k=\gamma  \) in equation (\ref{diq_su3})
and write the tensor product as:

\begin{equation}
\label{tens}
\epsilon ^{ijI}\epsilon ^{\alpha \beta I}=\sum _{i,j,\alpha ,\beta }(\delta _{i,\alpha }\delta _{j,\beta }-\delta _{i,\beta }\delta _{j,\alpha })
\end{equation}

We see that in this case the rotations in color and flavor space are no longer
independent but locked. Hence the quarks are in a color-flavor locked phase
if all three quark flavors participate at the formation of the diquark condensates.
\( \delta _{i,\alpha } \) is the unit matrix of \( SU(3)_{C\times F} \) in
which the matrices contain as columns the three flavors and as rows the three
colors. The Lagrangian is therefore invariant under a \( SU(3)_{C\times F} \)
transformation and consequently the \( SU(3) \) color and flavor symmetries
are reduced to an \( SU(3)_{C\times F} \) symmetry. For more consequences of
the appearance of the condensate for the symmetries we refer to the literature
\cite{sch99}. Here we will focus on a numerical evaluation of the size of the
condensates and the phase transitions at finite temperature and density.

\subsection{Thermodynamics}

As before in the case of the chiral phase transition we will evaluate all condensates
and the phase diagram by the evaluation of the thermodynamical potential.

We start by writing the Lagrangian in a more symmetric form, following Nambu
who developed this formalism for the classical superconductivity \cite{nam}.
For this purpose we rewrite the Lagrangian as a sum of the original Lagrangian
and its charge conjugate:

\begin{equation}
\LL _{Nambu}=\LL +\LL ^{C}
\end{equation}
Then the Lagrangian can be presented as a matrix:

\begin{equation}
\label{lanam_su3}
\LL _{Nambu}^{MF}=\left[ \begin{array}{cc}
\frac{1}{\sqrt{2}}\bar{q} & \frac{1}{\sqrt{2}}\bar{q}^{C}
\end{array}\right] \left[ \begin{array}{cc}
i\feyn {\partial }-M_{f} & \tilde{\Delta }^{k\gamma \dagger }\\
\tilde{\Delta }^{k\gamma } & -i\feyn {\partial }^{T}-M_{f}
\end{array}\right] \left[ \begin{array}{c}
\frac{1}{\sqrt{2}}q\\
\frac{1}{\sqrt{2}}q^{C}
\end{array}\right] +\LL _{cond}
\end{equation}
where we suppressed the indices for convenience and defined the term
\begin{equation}
\LL _{cond}=-2G_{S}(\alpha ^{2}+\beta ^{2}+\gamma ^{2})+4G_{D}\alpha \beta \gamma +\frac{|\Delta ^{k\gamma }|^{2}}{4G_{DIQ}}.
\end{equation}
 In order to calculate the thermodynamical potential in this notation
\[
\Omega =-\beta Tr[\ln exp(-\beta (\hat{H}_{Nambu}-\mu \hat{N}-\mu \hat{N}^{C}))]\]
we need the particle number operator and its charge conjugate:
\begin{equation}
\label{numb_p}
\hat{N}=\sum _{\vec{p},s,f,c}\left[ \hat{a}_{\vec{p},s}^{\dagger }\hat{a}_{\vec{p},s}-\hat{b}_{\vec{p},s}^{\dagger }\hat{b}_{\vec{p},s}\right] \quad \hat{N}^{C}=-\sum _{\vec{p},s,f,c}\left[ \hat{a}_{\vec{p},s}\hat{a}_{\vec{p},s}^{\dagger }-\hat{b}_{\vec{p},s}\hat{b}_{\vec{p},s}^{\dagger }\right] 
\end{equation}
where we suppressed the explicit dependence of the operators on flavor and color
degrees of freedom.

When calculating the Hamiltonian in the mean field approximation, one can see
that it is possible to separate \( \hat{H}-\mu \hat{N} \) into two parts, one
for the quarks (operators \( \hat{a} \) and \( \hat{a}^{\dagger } \)) and
another for the antiquarks (operators \( \hat{b} \) et \( \hat{b}^{\dagger } \)):
\begin{equation}
\hat{H}-\mu \hat{N}=(\hat{H}-\mu \hat{N})_{\hat{a}}+(\hat{H}-\mu \hat{N})_{\hat{b}}
\end{equation}
 These two parts yield the explicit expressions
\begin{equation}
\label{ha3_a}
(\hat{H}-\mu \hat{N})_{\hat{a}}=\sum _{\vec{p},s,f,c}\left[ \begin{array}{cc}
\hat{a}^{\dagger }_{\vec{p},s} & \hat{a}_{-\vec{p},-s}
\end{array}\right] \left[ \begin{array}{cc}
E_{\vec{p},f}-\mu _{f} & -\tilde{\Delta }^{k\gamma \dagger }N(p)\\
\tilde{\Delta }^{k\gamma }N(p) & -E_{\vec{p},f}+\mu _{f}
\end{array}\right] \left[ \begin{array}{c}
\hat{a}_{\vec{p},s}\\
\hat{a}^{\dagger }_{-\vec{p},-s}
\end{array}\right] +H_{cond}
\end{equation}
and 
\begin{equation}
\label{ha3_b}
(\hat{H}-\mu \hat{N})_{\hat{b}}=\sum _{\vec{p},s,f,c}\left[ \begin{array}{cc}
\hat{b}^{\dagger }_{\vec{p},s} & \hat{b}_{-\vec{p},-s}
\end{array}\right] \left[ \begin{array}{cc}
E_{\vec{p},f}+\mu _{f} & -\tilde{\Delta }^{k\gamma \dagger }N(p)\\
\tilde{\Delta }^{k\gamma }N(p) & -E_{\vec{p},f}-\mu _{f}
\end{array}\right] \left[ \begin{array}{c}
\hat{b}_{\vec{p},s}\\
\hat{b}^{\dagger }_{-\vec{p},-s}
\end{array}\right] +H_{cond.}
\end{equation}
 We denoted by \( H_{cond} \) the expression \( -V*\LL _{cond} \) and used
here the discrete summation over the momenta. The expressions have a defined
structure in flavor and color, the diagonal terms are diagonal in flavor and
color, the off-diagonal terms (\( \tilde{\Delta } \)) are antisymmetric in
color and flavor and 
\begin{eqnarray}
N(p)_{f_1,f_2}&=&\left( 1+\frac{p^2}{(E_{f_1}+m_{f_1})(E_{f_2}+m_{f_2})}\right)  \nonumber \\
&\times&\sqrt{\frac{(E_{f_1}+m_{f_1})(E_{f_2}+m_{f_2})}{{4 m_{f_1} m_{f_2}}}} \sqrt{\frac{m_{f_1} m_{f_2}}{E_{f_1} E_{f_2}}}         
\end{eqnarray}
This normalization factor is due to the fact that we deal with 
products of spinors for different species in the off diagonal terms, 
of course $N(p)=1$ when $f_1=f_2$. The explicit form of this matrix including
all flavor and color indices is displayed in appendix C.

In order to calculate the thermodynamical potential, we have
to diagonalize these expressions. This has to be done by means of a Bogolyubov
transformation which determines the energies of the quasi-particles and the
corresponding quasi-particle operators. From the discussion of the symmetry
of the diquark condensate we expect two quarks of different flavor and color
to form a diquark condensate whereas one quark of the third flavor is not involved 
in forming this condensate. This has to be seen in the quasi-particle 
energy and is confirmed
if we evaluate explicitly the quasi-particle energies as the eigenvalues of
the matrices. The diagonalized operators corresponding to \( (\hat{H}-\mu \hat{N})_{\hat{a}} \)
can be expressed in the form:

\[
(\hat{H}-\mu \hat{N})_{\hat{a}_{\Delta }}=\sum ^{3}_{i,\alpha =1}(E_{i\alpha }\hat{a}^{\dagger }_{\Delta i\alpha }\hat{a}_{\Delta i\alpha }+E_{i\alpha }'\hat{a}_{\Delta i\alpha }\hat{a}^{\dagger }_{\Delta i\alpha })\]
 where \( i \) runs over the flavors and \( \alpha  \) over the colors. \( \hat{a}_{\Delta i\alpha } \)
and \( \hat{a}^{\dagger }_{\Delta i\alpha } \) are annihilation and creation
operators for the quasi particles. 
:\begin{center}
\begin{tabular}{|c||c|c|} \hline
i &         $E_a,i$               & $g_i$ \\ \hline
1 & $\pm\sqrt{\Delta_{qq} + {E^-} ^2}$ & 3 \\ \hline
2& $\pm \frac{1}{2}( {Z} +E^--E^-_s)$& 2 \\ \hline
3& $\pm \frac{1}{2}( {Z} -E^-+E^-_s)$& 2 \\ \hline
4 & $\sqrt{Y-X}$ & 1 \\ \hline
5 & $\sqrt{Y+X}$ & 1 \\ \hline
\end{tabular}
\end{center}
where 
\begin{eqnarray}
E^- & = & E-\mu \\
E^-_s &= & E_s-\mu_s \\ 
Z &=& \sqrt{4 \Delta_{qs} N^2(p) +(E^-+E^-_s)^2} \\ 
Y &= &\frac{1}{2}\left(\Delta_{qq}^2 + 4 \Delta_{qs}^2 N^2(p) + {E^-} ^2 + {E^-_s} ^2 \right ) \\
X^2 &=& \frac{1}{4} \Bigl( \Delta_{qq}^2 + (8\Delta_{qs}^2 N^2(p) + (E^- + E_s^-)^2)(E^--E_s^-)^2 \nonumber \\
 & + & 2 \Delta_{qq}^2 \bigl(4 \Delta_{qs}^2 N^2(p)+(E^-+E_s^-)(E^--E_s^-) \bigr) \Bigr )
\end{eqnarray}
and $g_i$ is the degeneracy.
For the calculation of the thermodynamical potential it is not necessary to
know the exact form of the Bogolyubov transformation which relates the quasi
particle operators \( \hat{a}_{\Delta } \) with the original quark operators
\( \hat{a} \). The quasi-particles are still fermions, and that is all information
we need in order to evaluate the sum over the occupied states. It is just necessary
to assign the right energies to the operators. We evaluate the thermodynamical
potential for the case of two degenerated light quarks:

\begin{eqnarray}
\frac{\Omega }{V}  = \frac{\Omega _{0}}{V}-\frac{2}{\beta}\int ^{\Lambda }_{0}\frac{d\vec{p}}{(2 \pi)^3} &\Bigl \{ & 6 \ln [1 + \exp(-\beta E_{1-})] \nonumber \\
& +& 4 \ln [1 + \exp(-\beta E_{2-})] \nonumber \\
& +& 4 \ln [1 + \exp(-\beta E_{3-})] \nonumber \\
& +& 2 \ln [1 + \exp(-\beta E_{4-})] \nonumber \\
& +& 2 \ln [1 + \exp(-\beta E_{5-})] \nonumber \\
+3\beta E_{1-}+2\beta E_{2-}& + & 2\beta E_{3-}+\beta E_{4-}+\beta E_{5-} \Bigl \}
\end{eqnarray}
with
\begin{equation}
\frac{\Omega _{0}}{V}=4G_{S}(\alpha ^{2}+\beta ^{2}+\gamma ^{2})+\frac{2|\Delta _{qs}|^{2}+|\Delta _{qq}|^{2}}{2G_{DIQ}}
\end{equation}

This thermodynamical potential contains the (quark
and diquark) condensates as parameters. In order to evaluate them, we have to
minimize

\begin{equation}
\frac{\partial \Omega }{\partial \llt \bar{q}_{f}q_{f}\rrt }=0\qquad \frac{\partial \Omega }{\partial \Delta _{f_{1}f_{2}}}=0.
\end{equation}

 This minimization yields the gap equations for the quark condensates. For the
 SU(3) case the derivation is given in the appendix D.
 These equations are coupled, we have to solve them selfconsistently.
 The resulting $<qq>$ condensates may be found in appendix E. 

\subsection{Results at finite temperature and density}
 For this part we decide to take parameter in ref. \cite{Bentz01} :$m_{0,q}=5.96MeV$, $\Lambda=592.7 MeV$, $G_S=6.92 GeV^{-2}$, $G_{DIQ}/G_S =3./4.$ and $m_{0,s}=130.7 MeV$. We use the relation between the coupling 
constants, ( $G_{DIQ}=3 G_{S}/4$ ), given by the Fierz-transformation (see
appendix A), close to \cite{Bentz01} (0.73).

\begin{figure}
{\par\centering \resizebox*{0.8\textwidth}{!}{\includegraphics{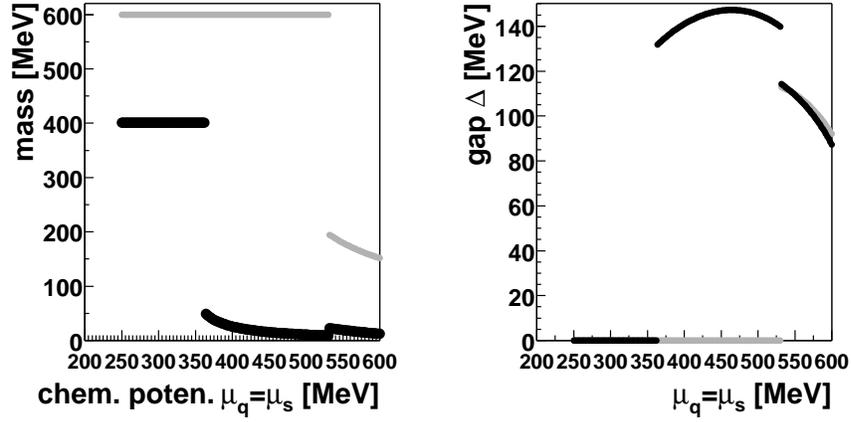}} \par}

\caption{\emph{The light (dark line) and strange (light line) quark masses and the diquark
condensates \protect\( \Delta _{qq}\protect \) (dark line) and \protect\( \Delta _{qs}\protect \)
(light line) as a function of the chemical potential \protect\( \mu _{q}=\mu _{s}\protect \)
at zero temperature.}\label{bild_muqs_mqs_dqs}}
\end{figure}

The condensates resp. masses at zero temperature as a function of the chemical
potential \( \mu _{q}=\mu _{s} \) are displayed in figure~\ref{bild_muqs_mqs_dqs}.
On the lhs of this figure we show the light and strange quark mass, on the rhs
the diquarks condensates. First we have to note that quark and diquark 
condensates compete with each other as they are formed by the same quarks. 
Temperature and density determine which condensate dominates.

When the chiral phase transition occurs (the quark condensate disappears), 
we observe for the light quarks that the superconducting phase transition takes
place and we have a diquark condensate. As the two transitions are related, 
they are of the same order. The same scenario repeats itself for the
strange diquark condensate at a higher chemical potential.
In the
figure we display only the solution which is the global minimum of the thermodynamical
potential.

\begin{figure}
{\par\centering \resizebox*{1\textwidth}{!}{\includegraphics{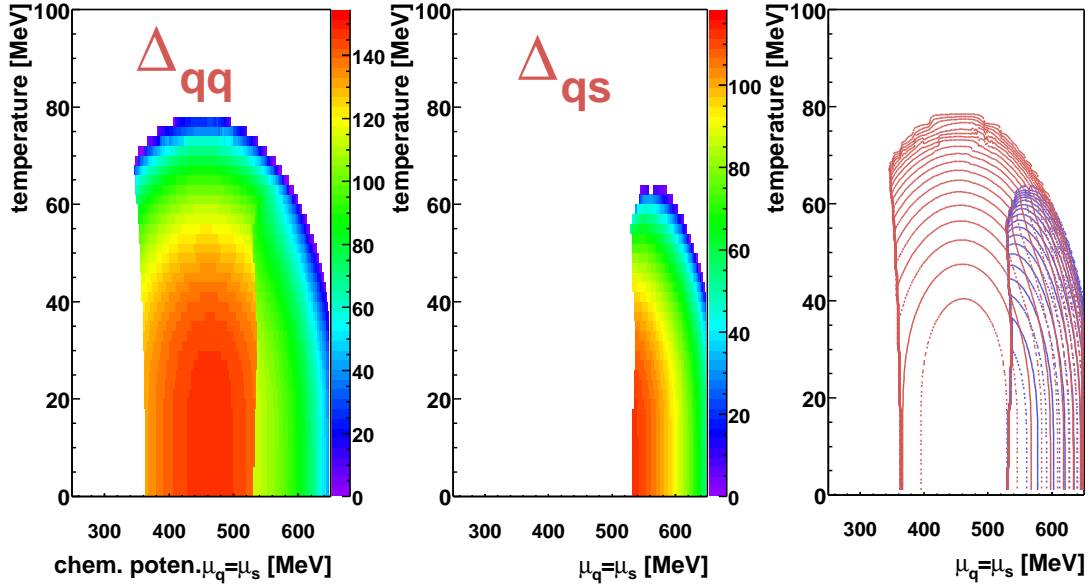}} \par}

\caption{\emph{Strength of the diquark condensates as a function of the chemical potential
\protect\( \mu _{q}=\mu _{s}\protect \) and the temperature. We show as color-levels
the strength of the condensates \protect\( \Delta _{qq}\protect \) (lhs), \protect\( \Delta _{qs}\protect \)
(middle) and superimpose both as a contour plot (rhs).} \label{bild_dqqdqs_mut}}
\end{figure}

At a quite low chemical potential (the light quarks have a very small mass,
the strange quarks are heavy) we have only the light diquark condensate, the
diquark condensate including strange quarks is almost zero as the strange quarks
show a strong quark condensate. Only when the strange quark condensate drops
and the mass of the strange quarks approaches its current mass, the strange
diquark condensate appears. We have here the coexistence of the light and strange
diquark condensate, this is the regime where the chiral symmetry is broken again
and color and flavor are locked. This happens at a quite high chemical potential,
the decreasing diquark condensate for even higher chemical potentials indicates
that we reach the limit of the model: we are too close to the cut-off. The phase
transitions concerning the strange quarks are quite close to the limits of the
models if we suppose the current mass of the strange quark of around \( 140\, MeV \).
We note that due to the relatively small difference between the quark masses,
both diquark condensates have approximately the same value, for the maximum
we get \( \Delta _{qq}\approx \Delta _{qs} \approx (120\, MeV)^{3} \). 
At zero temperature the chiral phase transition (where the quark condensates
disappear) and the superconducting phase transition (where the diquark condensates
appear) are strongly related in of our model. This changes at higher temperatures.
There the diquark condensates $\Delta_{qq}$ extends to smaller values of the 
chemical potential whereas we need higher densities in order to form a
$\Delta_{qs}$ diquark condensate. In addition the diquark condensate become
smaller with increasing temperature. This
is shown in figure~\ref{bild_dqqdqs_mut} where we plot the diquark condensates
as a function of temperature and chemical potential. For a given chemical potential
we observe - as in the classical superconductivity - a second order phase transition
as a function of the temperature.

\begin{figure}
{\par\centering \resizebox*{1\textwidth}{!}{\includegraphics{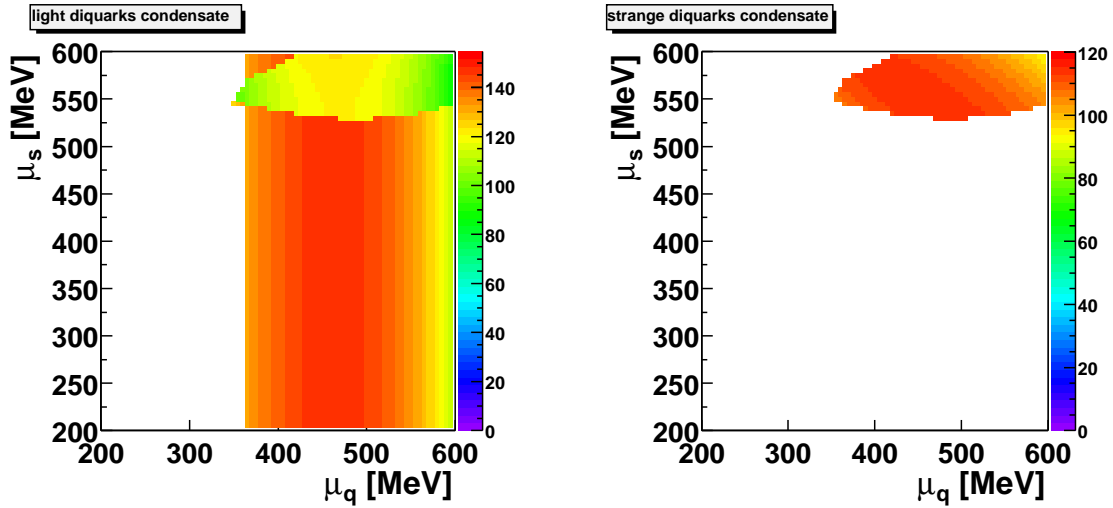}} \par}

\caption{\emph{As a function of the chemical potential \protect\( \mu _{q}\protect \)
and \protect\( \mu _{s}\protect \) we show the quark masses on the upper row
(light quarks lhs, strange quarks rhs) and the diquark condensates in the lower
row (\protect\( \Delta _{qq}\protect \) on the lhs and \protect\( \Delta _{qs}\protect \)
on the rhs)\label{bild_mqs_dqs_muq_mus}}}
\end{figure}

In a next step we consider the diquark condensates in the \( \mu _{q}-\mu _{s} \)
plane. As already mentioned, we expect the formation of a diquark condensate
only if there are quarks with similar Fermi momentum, independent of their mass.
Because $G_D$ is zero the disappearance of the quark condensates $<s\bar s>$ and
$< q \bar q>$  does not depend on the chemical potential of the other species.
There is one exception the creation of the strange diquark condensates lowers
the strange quark condensate and increases the light quark mass. 
In figure~\ref{bild_mqs_dqs_muq_mus} we plot the strength of the diquark 
condensates. Because both light quarks have the same chemical potential a 
diquark condensate \( \Delta _{qq} \) between the two different flavors occurs
whenever the light quark mass is small. 

The strange diquark condensate exists only in a band where the chemical potential
of the light and strange quarks are approximately equal. The slight deformation
of this band is due to the different current quark masses. The width of the
band is determined by the coupling strength: if the coupling in the diquark
sector is strong, the quarks can bind and form a condensate even if their chemical
potentials are quite different. For a small coupling strength, the chemical
potentials of the two quarks have to be (approximately, in case of different
quark masses) equal in order to form a diquark condensate.

\begin{figure}
{\par\centering \resizebox*{1\textwidth}{!}{\includegraphics{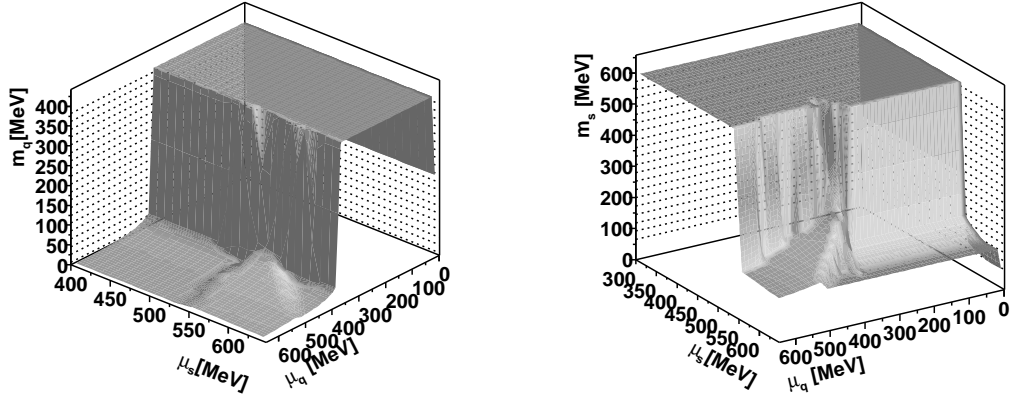}} \par}

\caption{\emph{As a function of the chemical potential $ \mu _{q}$ 
and \protect\( \mu _{s}\protect \) we show the quark masses 
(light quarks lhs, strange quarks rhs) to pointed out the effect of the diquark condensate on the quark condensate at T=1 MeV.}}
\label{mqms2}
\end{figure}

Now we should study the feedback of the formation of the diquark condensates 
on the quark condensates (or the mass). In \ref{mqms2} we display the masses of
the light and strange quarks as a function of $\mu_q$ and $\mu_s$,  
$\Delta_{qq}$ appears at the chiral phase transition, when the light quark
condensate disappears and it is formed by the free light quarks. This behavior
is almost independent of $\mu_s$. Only if $\Delta_{sq}$ becomes finite 
the lack of quarks for the quark condensate increases the light quark mass. 
The behavior  of $\Delta_{qs}$ is generic: When the diquark condensate 
$\Delta_{sq}$ is finite it takes quarks from the strange quark condensate
lowering the mass of the strange quark.
 

\section{Conclusions \label{chap_concl}}

In conclusion, we presented the phase diagram of the \( SU(3) \) flavor NJL
model extended to the diquark sector for a set of parameters which reproduces
meson masses and coupling constants. We find a rich structure of condensates
and regions where no condensate exists. The temperature and density dependence
of quark and diquark condensates is calculated in mean field approach by minimizing
the thermodynamical potential.

The order of chiral phase transition depends on the value of T and \( \mu  \)
where the phase transition occurs. At zero temperature the phase transition
is first order and at zero chemical potential we observe a cross over ( due
to finite current quark masses). Therefore there exists a tricritical point.
Normal nuclear matter density does exist only as a mixed phase of a dense quark
phase (where chiral symmetry is partially restored) and a very diluted quark
gas or the vacuum (where chiral symmetry is spontaneously broken). Finally we
extended the chiral phase transition to the plane of finite strange quark density,
relevant for the discussion of the diquark condensates.

Following the idea that the NJL model can be considered as an approximation
of the QCD Lagrangian we extend the NJL model by including an interaction in
the diquark channel. We find that this interaction gives rise to a diquark condensation
which is responsible for the formation of a superconducting gap. This condensation
occurs at low temperature and high density. As this gap is formed by two quarks
of different flavor, their chemical potential has to be close to each other
in order to allow for this condensation. In flavor \( SU_{F}(3) \) this condensate
breaks the \( SU_{F}(3)\times SU_{C}(3) \) down to a \( SU_{C\times F}(3) \)
a phenomenon which has already been observed in phase diagrams based on instanton
Lagrangians and has been dubbed color-flavor locking. We can conclude that two
quite differently motivated phenomenological approaches to the QCD Lagrangian
provide a very similar phase structure.

The diquark condensates do not exist at temperatures expected to be obtained
in relativistic heavy ion collisions. In neutron stars, which have a high density
and a very low temperature they could be of relevance.

\subsubsection*{Acknowledgments}

This work was supported by the Landesgraduiertenförderung 
Mecklenburg-Vorpommern.
One of us (J.A.) acknowledge an interesting discussion with K. Rajagopal.
We thank  A. W. Steiner and M. Prakash  of having pointed out an error in a formula of a previous version of this article.

\newpage
\appendix
\section{Fierz Transformation}
Following Ebert (\cite{ebe94}) we have the following relations in color and
flavor space:
\begin{eqnarray}
\delta_{ij}\delta_{kl}&=&\frac{1}{3}\delta_{ik}\delta_{lj}+\frac{1}{2}\sum_{a=1}^8 \lambda^a_{ik} \lambda_{lj} \;\;\;\;\;\;  (q\bar{q} \;\;\;\;\;channel)\\
\delta_{ij}\delta_{kl} &=& \frac{1}{2}\sum_{a=0}^{8}\lambda^a_{il} \lambda^a_{kj} \; \; \; \; \; \; (q q \; \; \; \; channel) \\
\sum_{a=1}^{8} \lambda^a_{ij}\lambda^a_{kl}&=& \frac{16}{9}\delta_{il}\delta_{kj}
-\frac{1}{3}\sum_{a=1}^{8}\lambda^a_{il}\lambda_{kj} \;\;\; \; \; \;  (q \bar{q} \; \; \; \; channel) \label{Fierzl1}\\
\sum_{a=1}^{8} \lambda^a_{ij}\lambda^a_{kl}&=& \frac{2}{3}\sum_{a=0,1,3,4,6,8}\lambda^a_{ik}\lambda^a_{lj}
-\frac{4}{3}\sum_{a=2,5,7}\lambda^a_{ik}\lambda^a_{lj}  \; \; \; \; \; \;  (q q \; \; \; \; channel )\label{Fierzl3}
\end{eqnarray}
\begin{equation}
\mathcal{L} = -g\sum_{a=1}^{8} (\bar{\psi} \gamma^C_{\mu} \lambda^a \psi)^2 
\end{equation}
This leads to the following relation between the differents coupling constant:
\begin{eqnarray}
q \bar{q} \; \; \; \; { \mathsf{channel}} \; \; \; G_{SCA} &=&\frac{8}{9}g \\ 
q q \; \; \; \; { \mathsf{channel}} \; \; \; G_{DIQ} &=&\frac{2}{3}g \\ 
\end{eqnarray}
\section{$<q\bar q>$ condensates}
\subsection{light quark condensate}
\begin{eqnarray}
\ll \bar{q}q \gg &=& -\frac{1}{2}\int \frac{p^2 dp}{2 \pi^2} \frac{M_q}{E_q}
\Bigl [3 \frac{E^-}{E^1_-} \nonumber \\ 
& + & f^\prime(E^2_-)\Bigl (\frac{1}{Z}\Bigl( 4 \ds^2 N(p) \dNdq \arrowvert_q +
E^-+E_s^-)+1 \Bigl) \nonumber \\
 & + & f^\prime(E^3_-) \Bigl (\frac{1}{Z}\Bigl( 4 \ds^2 N(p) \dNdq \arrowvert_q
 + E^-+E_s^-)-1 \Bigl) \nonumber \\
&+& \frac{f^\prime(E^4_-)}{2E^4_-} \Bigl [ 4 \ds^2 N(p)\dNdq \arrowvert_q+ E^- 
\nonumber\\ 
&\;& \; \; \; \; -\frac{1}{4X} \Bigl (  \Bigl [ (E^--E^-_s)(8
\ds^2N^2+(E^-+E_s^-)^2) \nonumber \\
&\;& \; \; \; \;  + (E^--E_s^-)^2(8 \ds^2 N(p) \dNdq \arrowvert_q +(E^-+E^-_s))
\nonumber\\
 &\;& \; \; \; \;   + \dq^2(8 \ds^2 N(p) \dNdq \arrowvert_q +2E^-) \Bigr ]
 \Bigr) \Bigr] \nonumber \\
& +& \frac{f^\prime(E^5_-)}{2E^5_-} \Bigl [ 4 \ds^2 N(p)\dNdq \arrowvert_q+ E^-
\nonumber\\ 
 &\;& \; \; \; \; +\frac{1}{4X} \Bigl ( \Bigl [ (E^--E^-_s)(8
 \ds^2N^2+(E^-+E_s^-)^2) \nonumber \\
& \;& \; \; \; \;+(E^--E_s^-)^2(8 \ds^2 N(p) \dNdq \arrowvert_q +(E^-+E^-_s))
\nonumber\\
&\;& \; \; \; \; + \dq^2(8 \ds^2 N(p) \dNdq \arrowvert_q +2E^-) \Bigr ] \Bigr)
\Bigr]  \nonumber \\ 
&+& (-) \Longrightarrow (+) \Bigr ]
\end{eqnarray}
\subsection{strange quark condensate}
\begin{eqnarray}
\ll \bar{s}s \gg &=&  -\int \frac{p^2 dp}{2 \pi^2} \frac{M_s}{E_s} \Bigl [
\frac{E^-}{E^1_-} \nonumber \\
&+&  f^\prime(E^2_-)\Bigl (\frac{1}{Z}( 4 \ds^2 N(p) \dNds \arrowvert_q +
E^-+E_s^-)-1 \Bigl) \nonumber \\
&+&  f^\prime(E^3_-)\Bigl (\frac{1}{Z}( 4 \ds^2 N(p) \dNds \arrowvert_q +
E^-+E_s^-)+1 \Bigl) \nonumber \\
&+&\frac{f^\prime(E^4_-)}{2E^4_-} \Bigl [4 \ds^2 N(p)\dNds \arrowvert_s  +E^-_s
\nonumber \\
&\;& \; \; \; \; - \frac{1}{4X} \Bigl (   \Bigl [ -(E^--E^-_s)(8
\ds^2N^2+(E^-+E_s^-)^2)\nonumber \\
&\;& \; \; \; \;  + (E^--E_s^-)^2(8 \ds^2 N(p) \dNds \arrowvert_s
+(E^-+E^-_s))\nonumber \\
&\;& \; \; \; \; + \dq^2(8 \ds^2 N(p) \dNds \arrowvert_s -2E^-_s) \Bigr ] \Bigr)
\Bigr] \\
 &+& \frac{f^\prime(E^5_-)}{2E^5_-} \Bigl [4 \ds^2 N(p)\dNds \arrowvert_s 
 +E^-_s \nonumber \\
&\;& \; \; \; \; + \frac{1}{4X} \Bigl (   \Bigl [ -(E^--E^-_s)(8
\ds^2N^2+(E^-+E_s^-)^2)\nonumber \\
&\;& \; \; \; \;  + (E^--E_s^-)^2(8 \ds^2 N(p) \dNds \arrowvert_s
+(E^-+E^-_s))\nonumber \\
&\;& \; \; \; \; + \dq^2(8 \ds^2 N(p) \dNds \arrowvert_s -2E^-_s) \Bigr]  \Bigr)
\Bigr] \nonumber \\
&+& (-) \Longrightarrow (+) \Bigr ]
\end{eqnarray}
\section{Matrix}
The total matrix can be separated into 4 submatrices
\begin{center}
\begin{tabular}{|c||c|c|} \hline
 & $a$ & $a^{\dagger}$ \\ \hline
$a^{\dagger}$ & A  & C \\ \hline
a             & -$C^\dagger$ & B  \\ \hline 
\end{tabular}
\end{center}
These submatrices are given by:
\begin{center}
A=
\begin{tabular}{|c||c|c|c|c|c|c|c|c|c|} \hline
 & $u_R$ & $u_G$ & $u_B$ &  $d_R$ & $d_G$ & $d_B$ &  $s_R$ & $s_G$ & $ss_B$ \\ \hline \hline
$\uc_R$ &$E^-$&0&0&0&0&0&0&0&0 \\ \hline
$\uc_G$ &0&$E^-$&0&0&0&0&0&0&0\\ \hline
$\uc_B$ &0&0&$E^-$&0&0&0&0&0&0\\ \hline
$\dc_R$ &0&0&0&$E^-$&0&0&0&0&0\\ \hline
$\dc_G$ &0&0&0&0&$E^-$&0&0&0&0\\ \hline
$\dc_B$ &0&0&0&0&0&$E^-$&0&0&0\\ \hline
$\Sc_R$ &0&0&0&0&0&0&$E_s^-$&0&0\\ \hline
$\Sc_G$ &0&0&0&0&0&0&0&$E^-_s$&0\\ \hline
$\Sc_B$ &0&0&0&0&0&0&0&0&$E_s^-$\\ \hline
\end{tabular}
\end{center}
\begin{center}
B=
\begin{tabular}{|c||c|c|c|c|c|c|c|c|c|} \hline
   & $\uc_R$ & $\uc_G$ & $\uc_B$ &  $\dc_R$ & $\dc_G$ & $\dc_B$ &  $\Sc_R$ & $\Sc_G$ & $\Sc_B$ \\ \hline \hline
$u_R$ &$E^+$&0&0&0&0&0&0&0&0 \\ \hline
$u_G$ &0&$E^+$&0&0&0&0&0&0&0\\ \hline
$u_B$ &0&0&$E^+$&0&0&0&0&0&0\\ \hline
$d_R$ &0&0&0&$E^+$&0&0&0&0&0\\ \hline
$d_G$ &0&0&0&0&$E^+$&0&0&0&0\\ \hline
$d_B$ &0&0&0&0&0&$E^+$&0&0&0\\ \hline
$s_R$ &0&0&0&0&0&0&$E_s^+$&0&0\\ \hline
$s_G$ &0&0&0&0&0&0&0&$E^+_s$&0\\ \hline
$s_B$ &0&0&0&0&0&0&0&0&$E_s^+$\\ \hline
\end{tabular}
\end{center}
\begin{center}
C=
\begin{tabular}{|c||c|c|c|c|c|c|c|c|c|} \hline
   & $\uc_R$ & $\uc_G$ & $\uc_B$ &  $\dc_R$ & $\dc_G$ & $\dc_B$ &  $\Sc_R$ & $\Sc_G$ & $\Sc_B$ \\ \hline \hline
$\uc_R$ &0&0&0&0&$\dq$&0&0&0&$\ds$ \\ \hline
$\uc_G$ &0&0&0&-$\dq$&0&0&0&0&0\\ \hline
$\uc_B$ &0&0&0&0&0&0&-$\ds$&0&0\\ \hline
$\dc_R$ &0&$-\dq$&0&0&0&0&0&0&0\\ \hline
$\dc_G$ &$\dq$&0&0&0&0&0&0&0&$\ds$\\ \hline
$\dc_B$ &0&0&0&0&0&0&0&-$\ds$&0\\ \hline
$\Sc_R$ &0&0&-$\ds$&0&0&0&0&0&0\\ \hline
$\Sc_G$ &0&0&0&0&0&-$\ds$&0&0&0\\ \hline
$\Sc_B$ &$\ds$&0&0&0&$-\ds$&0&0&0&0\\ \hline
\end{tabular}
\end{center}

\section{Thermodynamical potential}
\subsection{formal derivation of the thermodynamical potential}
\begin{eqnarray}
\dxda{\Omega}=\dxda{\Omega_0}- \frac{1}{\beta} \int\frac{d^3\vec{p}}{(2\pi)^3} &\Bigl[&-6 \beta \dxda{E^1_-}f(E^1_-) \nonumber\\ 
&-&4 \beta \dxda{E^2_-}f(E^2_-) \nonumber \\
&-&4 \beta \dxda{E^3_-}f(E^3_-) \nonumber \\
&-&2 \beta \dxda{E^4_-}f(E^4_-) \nonumber \\
&-&2 \beta \dxda{E^5_-}f(E^5_-) \nonumber \\
+ 3\beta \dxda{E^1_-} + 2\beta \dxda{E^2_-} &+&2\beta \dxda{E^3_-} + \beta \dxda{E^4_-} +  \beta \dxda{E^5_-} \nonumber \\
&\;& (-) \rightarrow (+) \Bigr ]
\end{eqnarray}
\begin{eqnarray}
\dxda{\Omega}=\dxda{\Omega_0}-2\int\frac{d^3\vec{p}}{(4\pi)^3} &\Bigl[& 3\dxda{E^1_-} f^\prime(E^1_-)  + 2\dxda{E^2_-} f^\prime(E^2_-)\nonumber \\
 &+&  2\dxda{E^3_-} f^\prime(E^3_-) +\dxda{E^4_-} f^\prime(E^4_-)+\dxda{E^5_-} f^\prime(E^5_-) \Bigr]
\end{eqnarray}
where $f^\prime(x)=1-2f(x)$
\subsection{N(p) derivatives}
\begin{equation}
N(p)=\left( 1+\frac{p^2}{(E+m)(E_{s}+m_{s})}\right)   \sqrt{\frac{(E+m)(E_s+m_s)}{{4 m m_s}}} \sqrt{\frac{m m_s}{E E_s}}         
\end{equation}
Then we write :
\begin{eqnarray}
U &=&\left( 1+\frac{p^2}{(E+m)(E_{s}+m_{s})}\right)\\
V &=&\sqrt{\frac{m m_s}{E E_s}} \\ 
W &=& \sqrt{\frac{(E+m)(E_s+m_s)}{{4 m m_s}}} 
\end{eqnarray}
\begin{eqnarray}
\frac{\partial W}{\partial \alpha} = \frac{1}{2W} \frac{(E_q+m_q)(E_s+m_s)}{4 m_q^2 m_s^2} &\Bigl[& \dEda (\frac{m_s}{m_q}(m_q-E_q)) \\ 
&+&  \dEsda  (\frac{m_q}{m_s}(m_s-E_s)) \Bigr ] 
\end{eqnarray}
\begin{equation}
\frac{\partial W}{\partial \alpha}  = \dxda{E^-_q}\frac{\partial W}{\partial \alpha} \arrowvert_q + \dxda{E^-_s}\frac{\partial W}{\partial \alpha} \arrowvert_s
\end{equation}
\begin{eqnarray}
\frac{\partial V}{\partial \alpha} &=& \frac{1}{2V}\Bigl[ \dEda (\frac{m_s}{E_s m_q}-\frac{m_s m_q}{E_q^2 E_s}) +  \dEsda (\frac{m_q}{E_q m_s}-\frac{m_s m_q}{E_s^2 E_q})  \Bigr ]  \nonumber \\
\frac{\partial V}{\partial \alpha} &=&\dxda{E^-_q}\frac{\partial V}{\partial \alpha} \arrowvert_q + \dxda{E^-_s}\frac{\partial V}{\partial \alpha} \arrowvert_s
\end{eqnarray}
\begin{eqnarray}
\frac{\partial U}{\partial \alpha} &=&  \frac{-p^2}{(E_q+m_q)(E_s+m_s)} \Bigl[ \dEda \frac{1}{m_q}+  \dEsda  \frac{1}{m_s} \Bigr ]   \nonumber \\
\frac{\partial U}{\partial \alpha} &=&\frac{\partial U}{\partial \alpha} \arrowvert_q + \frac{\partial U}{\partial \alpha} \arrowvert_s
\end{eqnarray}
So ,
\begin{eqnarray}
\dNda \arrowvert_q &=&\frac{\partial U}{\partial \alpha} \arrowvert_q VW + U\frac{\partial V}{\partial \alpha}\arrowvert_q W+ UV \frac{\partial W}{\partial \alpha} \arrowvert_q \\
\dNda \arrowvert_s &=&\frac{\partial U}{\partial \alpha} \arrowvert_s VW + U\frac{\partial V}{\partial \alpha}\arrowvert_s W + UV \frac{\partial W}{\partial \alpha} \arrowvert_s
\end{eqnarray}
\section{<qq> condensates}
\subsection{light diquarks condensate}
\begin{eqnarray}
 \dq &=&  \frac{G_{diq}}{\pi^2} \int p^2 dp \Bigl [ \frac{3 \dq}{E^1_-} f^\prime(E^1_-)  \nonumber \\
&+& \frac{f^\prime(E^4_-)}{2E^4_-} \Bigl ( \dq-\frac{1}{2X}(\dq^3+\dq(4\ds^2(p)+{E^-}^2-{E^-}^2)) \Bigr)  \nonumber \\
&+& \frac{f^\prime({E^5}_-)}{2E^5_-} \Bigl ( \dq+\frac{1}{2X}(\dq^3+\dq(4\ds^2(p)+{E^-}^2-{E^-}^2)) \Bigr)  \nonumber \\
&+& (-) \Longrightarrow (+) \Bigl ] \\
\end{eqnarray}
\subsection{strange diquarks condensate}
\begin{eqnarray}
 \ds &=&  \frac{G_{diq}}{2 \pi^2} \int p^2 dp \Bigl [ \frac{4 \ds N^2(p)}{Z} f^\prime(E^2_-) + \frac{4 \ds N^2(p)}{Z} f^\prime(E^3_-)\nonumber \\
&+& \frac{f^\prime(E^4_-)}{2E^4_-} \Bigl (4\ds N^2(p) - \frac{4}{2X}\Bigl((E^--E^-_s)^2(\ds N^2(p))+\dq^2\ds N^2(p) \Bigr)\Bigr) \nonumber \\
&+& \frac{f^\prime(E^5_-)}{2E^5_-} \Bigl (4\ds N^2(p) +  \frac{4}{2X}\Bigl((E^--E^-_s)^2(\ds N^2(p))+\dq^2\ds N^2(p) \Bigr)\Bigr)  \nonumber \\
&+& (-) \Longrightarrow (+) \Bigl ] \\
\end{eqnarray}








\end{document}